# BRIEF ANALYSIS OF VORTICITY IN SOIL HYDRODYNAMICS

J. J. Nader

José Jorge Nader

Department of Structural and Geotechnical Engineering

Polytechnic School, University of São Paulo

05508-900, São Paulo, Brazil, e-mail: jjnader@usp.br, fax: 55-11-3091-5181

**Abstract**

This note discusses basic properties of vorticity in groundwater flow theory. An evolution equation for the vorticity vector is derived to demonstrate that, when present, vorticity decreases very rapidly. In addition, it is shown how vorticity affects, though very little, the hydraulic head directional variation in the vicinity of a point.

The classic theory of water motion in homogeneous, isotropic soils assumes that the velocity vector is the gradient of a potential and is therefore restricted to irrotational flows [1, 2, 3]. Differently, by starting from the general equation of motion, this note examines elementary properties of flows with non-zero vorticity. First, the evolution of the vorticity vector (the curl of the velocity) will be studied; the results will justify, from a theoretical perspective, the practical value of the traditional hypothesis of irrotationality. Moreover, it will be shown how vorticity affects the directional variation of hydraulic head in the neighbourhood of a point, thereby disturbing slightly the typical flow net configuration.

The analysis begins with the differential equation of motion: $-\text{grad}(p + \gamma z) - n\gamma \mathbf{v}/k = \rho \mathbf{a}$, relating the pressure $p$, the velocity $\mathbf{v}$, the acceleration $\mathbf{a}$ and a vertical coordinate $z$ pointing upwards. The constants $\rho$, $\gamma$, $n$ and $k$ denote the water density, the water unit weight, the porosity and the permeability coefficient, respectively. The term $-n\gamma\mathbf{v}/k$ represents the resistance force per unit volume applied by the grains on the moving water.

The vorticity vector, $\boldsymbol{\omega} = \operatorname{curl}\mathbf{v}$, is introduced in the discussion by means of the Lagrange formula for the acceleration [4]: $\mathbf{a} = \partial\mathbf{v}/\partial t + \operatorname{grad}(v^2)/2 + \boldsymbol{\omega}\times\mathbf{v}$ ($v$ denotes the norm of $\mathbf{v}$). The equation of motion, after introduction of $\mathbf{a}$ as given above and of the total hydraulic head, $h = p/\gamma + z + v^2/2g$ ($g$ is the gravitational acceleration), becomes:

$$\operatorname{grad} h = -\frac{n}{k}\mathbf{v} - \frac{1}{g}\left(\frac{\partial\mathbf{v}}{\partial t} + \boldsymbol{\omega}\times\mathbf{v}\right). \tag{1}$$

Note in passing that, for steady irrotational flows ($\partial\mathbf{v}/\partial t = \mathbf{0}$; $\boldsymbol{\omega} = \mathbf{0}$), eq. 1 reduces to Darcy's law (version II of Darcy's law in [5]). By taking the curl of both sides of eq. 1, $h$ is eliminated and an evolution equation for $\boldsymbol{\omega}$ is obtained:

$$\dot{\boldsymbol{\omega}} = (\operatorname{grad}\mathbf{v})\boldsymbol{\omega} - \frac{ng}{k}\boldsymbol{\omega}, \tag{2}$$

where $\dot{\boldsymbol{\omega}} = \partial\boldsymbol{\omega}/\partial t + (\operatorname{grad}\boldsymbol{\omega})\mathbf{v}$ is the material time derivative of $\boldsymbol{\omega}$, which measures the rate of change of the vorticity of a water particle. To arrive at eq. 2 the following relations have been employed: $\operatorname{div}\mathbf{v} = 0$ (equation of continuity for a constant porosity soil), together with the identities $\operatorname{div}\boldsymbol{\omega} = 0$ and $\operatorname{curl}(\boldsymbol{\omega}\times\mathbf{v}) = (\operatorname{div}\mathbf{v})\boldsymbol{\omega} - (\operatorname{div}\boldsymbol{\omega})\mathbf{v} + (\operatorname{grad}\boldsymbol{\omega})\mathbf{v} - (\operatorname{grad}\mathbf{v})\boldsymbol{\omega}$. For the sake of comparison with classic fluid mechanics, recall that in the theory of ideal fluids the term $-ng\boldsymbol{\omega}/k$ is absent, whereas in the theory of Newtonian viscous fluids it is replaced by the product of the kinematic viscosity and the Laplacian of the velocity; the term $(\operatorname{grad}\mathbf{v})\boldsymbol{\omega}$ is present in both theories [4, 6].

In plane flows $\boldsymbol{\omega}$ is orthogonal to the flow plane, so $(\operatorname{grad}\mathbf{v})\boldsymbol{\omega} = \mathbf{0}$. Therefore, $\dot{\omega} = -ng\omega/k$ and the vorticity of a particle falls exponentially in time ($\omega$ denotes the norm of $\boldsymbol{\omega}$). More explicitly, after integration one obtains $\omega_m(\mathbf{p},t) = \omega_0(\mathbf{p})\exp(-ngt/k)$ for the Langrangian (or material) description of the vorticity of a particle $\mathbf{p}$. In typical situations the decrease is very fast: if, for instance, $n$=0.3, $g$=9.81 m/s$^2$ and $k=10^{-5}$ m/s, then $ng/k \cong 2.9 \times 10^5 \, s^{-1}$. Note that, the smaller the permeability coefficient, the faster is the reduction.

In general three dimensional flows the interpretation of the term $-ng\boldsymbol{\omega}/k$ in eq. 2 is essentially the same: it represents a decrease in time of the magnitude of the vorticity of a particle proportional to the current vorticity (exponential decrease) without any directional change. The term $(\text{grad}\,\mathbf{v})\boldsymbol{\omega}$, in turn, affects $\boldsymbol{\omega}$ in a complex way; it may influence both its magnitude and direction [7]. In soil mechanics, however, since the components of the tensor $\text{grad}\,\mathbf{v}$ (i.e., the spatial partial derivatives of the velocity components, $\partial v_i/\partial x_j$) are typically far less, in absolute value, than $ng/k$, the magnitude of the term $(\text{grad}\,\mathbf{v})\boldsymbol{\omega}$ is usually very small in comparison with the dominant term $-ng\boldsymbol{\omega}/k$.

Thus, when present, vorticity decreases very rapidly along flow lines. Another important conclusion can be drawn immediately from eq. 2: if the vorticity of a particle is zero at a certain instant, then it is zero all the time. These strong results support the traditional assumption of irrotationality made in the classic theory of groundwater motion.

Finally, the influence of $\boldsymbol{\omega}$ on the directional variation of $h$ will be addressed. In the theory of steady irrotational flows in soils, the direction of greatest decrease in $h$ is the direction of the velocity and, therefore, level surfaces of $h$ (equipotential surfaces) are orthogonal to flow lines. The presence of vorticity affects this picture.

At a point of the flow domain, let $\mathbf{u}$, $\boldsymbol{\tau}$ and $\mathbf{e}$ be unit vectors in the directions of $\boldsymbol{\omega}$, $\mathbf{v}$ and $\boldsymbol{\omega} \times \mathbf{v}$, respectively (suppose $\boldsymbol{\omega} \times \mathbf{v} \neq \mathbf{0}$; the case $\boldsymbol{\omega} \times \mathbf{v} = \mathbf{0}$ is trivial). Thus, $\boldsymbol{\omega} = \omega\mathbf{u}$, $\mathbf{v} = v\boldsymbol{\tau}$ and $\boldsymbol{\omega} \times \mathbf{v} = \omega v \sin\theta \mathbf{e}$, where $\theta$ is the angle between $\boldsymbol{\omega}$ and $\mathbf{v}$. For steady flows, eq. 1 reduces to $\text{grad}\,h = -nv\boldsymbol{\tau}/k - \omega v \sin\theta \mathbf{e}/g$. Hence, the direction of greatest decrease in $h$, at a given point, is that of $\boldsymbol{\tau} + k\omega \sin\theta \mathbf{e}/ng$ (orthogonal to the level surface of $h$ passing through that point); the amount by which it deviates from the flow direction is indicated by the number $k\omega \text{sen}\theta/ng$ (usually very small). On the other hand, along a flow line, the decrease in $h$ per unit length is given by $\partial h/\partial \boldsymbol{\tau} = \text{grad}\,h.\boldsymbol{\tau} = -nv/k$ (the derivative of $h$ in the direction of $\boldsymbol{\tau}$), which is proportional to $v$, inversely proportional to $k$ and is independent of $\boldsymbol{\omega}$ (version III of Darcy's law in [5]). In contrast with what happens in irrotational flows, here, perpendicularly to flow lines, the derivative is not zero in general: it is zero only in the direction orthogonal to the plane defined by $\boldsymbol{\tau}$ and $\mathbf{e}$; in the direction of $\mathbf{e}$, $\partial h/\partial \mathbf{e} = -\omega v \sin\theta / g$ (independent of $k$).

The situation is easier to understand in the case of plane flows ($\theta = \pi/2$, $\mathbf{u}$ is orthogonal to the flow plane, $\boldsymbol{\tau}$ and $\mathbf{e}$ are parallel to the flow plane). Now, $\mathrm{grad}\, h = -nv\boldsymbol{\tau}/k - \omega v \mathbf{e}/g$. In the flow plane, at a given point, consider the vector basis $(\boldsymbol{\tau},\mathbf{e})$ and the polar angle $\alpha$ measured anticlockwise from $\boldsymbol{\tau}$. The direction of maximum change of $h$ is given by $\tan\alpha_M = k\omega/ng$ and that of zero variation by $\tan\alpha_0 = -ng/k\omega$. These directions deviate more from the directions of $\boldsymbol{\tau}$ and $\mathbf{e}$, respectively, for higher values of the product $\omega k$. For typical values, as those adopted before ($ng/k \cong 2.9 \mathrm{x} 10^5 s^{-1}$), the deviation is negligible.

## REFERENCES

1. Polubarinova-Kochina, P.Y.: Theory of Groundwater Movement. Princeton University Press, Princeton (1962).

2. Scheidegger, A.E.: The Physics of Flow through Porous Media. University of Toronto Press (1974).

3. Harr, M.E.: Groundwater and Seepage. Dover (1991).

4. Truesdell C.A., Toupin R.A.: The Classical Field Theories. In: Handbuch der Physik III/1, Springer (1960).

5. Nader, J.J.: Darcy´s law and the differential equation of motion. Géotechnique 59(6), 551-552 (2009).

6. Meyer, R.E. Introduction to Mathematical Fluid Mechanics. Dover (2010).

7. Majda, A.J., Bertozzi, A.L. Vorticity and Incompressible Flow. Cambridge University Press (2002).